\documentclass[structabstract]{aa}  
\usepackage{graphicx}
\usepackage{txfonts}
\usepackage{natbib}
\begin{document}
   \title{Water vapour at high redshift: Arecibo monitoring of the megamaser in MG~J0414+0534}

   \subtitle{}

   \author{P. Castangia
          \inst{1}
          \and
          C. M. V. Impellizzeri
          \inst{2}
          \and
          J. P. McKean
          \inst{3}
          \and
          C. Henkel
          \inst{4}
          \and
          A. Brunthaler
          \inst{4}
          \and
          A. L. Roy
          \inst{4}
          \and
          O. Wucknitz
          \inst{5}
          \and
          J. Ott
          \inst{6}
          \and
          E. Momjian
          \inst{6}
          }

   \institute{INAF-Osservatorio Astronomico di Cagliari,
              Loc. Poggio dei Pini, Strada 54, I-09012 Capoterra (CA), Italy\\
              \email{pcastang@oa-cagliari.inaf.it}
         \and
             National Radio Astronomy Observatory,
             520 Edgemont Road, Charllottesville, VA 22903, USA\\
             \email{vimpelli@nrao.edu}
         \and 
             ASTRON,
             Oude Hoogeveensedijk 4, 7991 PD Dwingeloo, the Netherlands\\
         \and
             Max-Planck-Institut f\"ur Radioastronomie,
             Auf dem H\"ugel 69, D-53121 Bonn, Germany\\
         \and
             Argelander-Institut f¨ur Astronomie, 
             Auf dem H\"ugel 71, D-53121 Bonn, Germany\\
         \and
            National Radio Astronomy Observatory,
            P.O. Box O, Socorro, NM 87801, USA
             }

   \date{Received ; accepted}

 
  \abstract
   {}
   {The study of water masers at cosmological distances would allow us to investigate the parsec-scale environment around powerful radio sources, to probe the physical conditions of the molecular gas in the inner parsecs of quasars, and to estimate their nuclear engine masses in the early universe. To derive this information, the nature of the maser source, jet or disk-maser, needs to be assessed through a detailed investigation of the observational characteristics of the line emission.}
   {We monitored the maser line in the lensed quasar MG~J0414+0534 at $z = 2.64$ with the 300-m Arecibo telescope for $\sim$15 months to detect possible additional maser components and to measure a potential velocity drift of the lines. In addition, we follow the maser and continuum emissions to reveal significant variations in their flux density and to determine correlation or time-lag, if any, between them.}
   {The main maser line profile is complex and can be resolved into a number of broad features with line widths of 30--160\,km\,s$^{-1}$. A new maser component was tentatively detected in October 2008 that is redshifted by 470\,km\,s$^{-1}$ w.r.t the systemic velocity of the quasar. The line width of the main maser feature increased by a factor of two between the Effelsberg and EVLA observations reported by Impellizzeri et al. (2008) and the first epoch of the Arecibo monitoring campaign. After correcting for the lens magnification, we find that the total H$_2$O isotropic luminosity of the maser in MG~J0414+0534 is now $\sim$30,000\,L$_{\odot}$, making this source the most luminous ever discovered. Both the main line peak and continuum flux densities are surprisingly stable throughout the period of the observations. The integrated flux density shows instead significant variations on monthly time scales, possibly due to changes in the individual velocity components. We place an upper limit on the velocity drift of the peak of the line emission of 2\,km\,s$^{-1}$\,yr$^{-1}$.}
   {The large line width of the main maser line and the absence of a clear triple-peak pattern in the maser spectrum of MG~J0414+0534 favors the jet-maser scenario. However, the stability of the line and continuum emission, and the presence of the tentative new maser component, potentially identified as a high-velocity feature of a rotating disk, seems partly going against this interpretation. Sensitive monitoring on a longer time-scale and VLBI observations are mandatory to draw a definite conclusion.}


   \maketitle
%

\section{Introduction}

The 22.2\,GHz radio emission from luminous extragalactic H$_2$O masers originates in dense ($10^7$\,$<$\,$n({\rm H_2})$\,$<$\,$10^{11}$\,cm$^{-3}$) and warm ($T$\,$>$\,300\,K) gas clouds within a few parsecs from the nuclear engines of their parent galaxies. These masers trace circumnuclear accretion disks (``disk-masers'', e.\,g. UGC3789; \citealt{reid09}), the inner parts of relativistic jets (``jet-masers'', e.\,g. Mrk~348; \citealt{peck03}) or nuclear outflows (Circinus; \citealt{greenhill03}), that are associated with active galactic nuclei (AGN). In contrast to optical and ultraviolet radiation, the radio photons can penetrate the enormous column densities of gas and dust that often obscure the line of sight to the nucleus. This, together with the high brightness temperature and small size of the maser spots, makes H$_2$O emission a suitable tool to investigate the geometry, kinematics, and excitation conditions of the gas in the immediate vicinity of supermassive black holes. Very Long Baseline Interferometry (VLBI) studies of water maser sources, complemented by single-dish monitoring, are a unique instrument to map accretion disks and to estimate the enclosed masses (e.\,g. \citealt{braatz2010}; \citealt{kuo2011}), as well as to determine the shock speeds and densities of radio jets \citep{peck03}. 

To date, most such studies have targeted radio quiet AGN in the local Universe. Indeed, the majority of the known extragalactic water masers have been found in  Seyfert 2 or LINER galaxies at $z<0.06$. However, the discovery of a water maser in a type 2 quasar at $z=0.66$ \citep{barvainis05} demonstrated that H$_2$O masers can also be detected at higher redshifts. The discovery of water masers at cosmological distances ($z>1.5$) would allow us to study the parsec-scale environment around powerful radio sources, to investigate the physical conditions of the molecular gas in the inner parsecs of quasars, and to measure their black-hole masses not only in the local but also in the early universe. We have recently performed a survey of gravitationally lensed quasars with the Effelsberg radio telescope to find water masers at cosmological redshifts (\citealt{impellizzeri08}; \citealt{mckean2010}). By observing gravitational lens systems we use the lens as a `cosmic telescope' to probe a luminosity regime that is otherwise not reachable with current instrumentation. 
Our first confirmed high redshift water maser was found toward the lensed quasar MG~J0414+0534 at $z=2.64$, which is by far the most distant object known to host water maser emission \citep{impellizzeri08}. The previously reported (unlensed) H$_2$O apparent isotropic luminosity of $\sim$10,000\,L$_{\odot}$ places the maser in MG~J0414+0534 among the most luminous water masers ever detected and suggests that the emission is associated with the AGN of the quasar. Although the characteristics of the spectrum seem to favour an association with the radio jet rather than with an accretion disk, the origin of the H$_2$O emission could not be conclusively determined from the Effelsberg and EVLA data alone. 

In this paper we report the results from 15 months of monitoring of the redshifted 6\,GHz radio continuum and line emission in MG~J0414+0534 with the 300-m Arecibo telescope. We monitored the line with a large bandwidth to potentially detect additional maser components and to constrain a possible velocity drift of the lines. Furthermore, we monitored the line to reveal possible variations in the maser flux density, determine if a correlation exists between the maser and the continuum flux density and whether there is a time-lag between them. Throughout the paper we adopt a cosmology with $\Omega_{\rm M} =0.3$, $\Omega_{\rm \Lambda} =0.7$ and $H_0 = 70$\,km\,s$^{-1}$\,Mpc$^{-1}$. 
      

\section{Observations and data reduction}\label{sect:obs}

The water maser line from the gravitationally lensed quasar MG~J0414+0534 was monitored with the Arecibo telescope between October 2008 and January 2010, at $\sim$6 week intervals, for a total of 11 epochs (see Table~\ref{table:obs}). 

We observed the 6$_{16}$--5$_{23}$ transition of ortho-H$_2$O (rest frequency 22.235\,GHz) using the C-high receiver when available and the standard C-band receiver otherwise. Both receivers provide dual linear polarization. For most of the observations (8 out of 11), we employed the Wideband Arecibo Pulsar Processor (WAPP) spectrometer in single board mode, which provides up to four independently tunable sub-correlators. We used two of the four WAPP sub-correlators, each with a 100\,MHz bandwidth centered at the redshifted frequency of the line (6.110\,GHz at $z$\,=\,2.639) and with a single polarization. In three epochs, August 2009, November 2009, and January 2010, we used the WAPP in dual board mode. This mode provides eight independent sub-correlators, each of 100\,MHz bandwidth, which can be centered at different frequencies within an instantaneous interval of 1\,GHz. We utilized the sub-correlators to simultaneously observe the water maser line and other redshifted molecular transitions including four ammonia inversion lines, NH$_3$ (1, 1), (2, 2), (3, 3), and (6, 6), and five excited OH and CH$_3$OH transitions (see Table~\ref{table:trans}). With nine-level quantization and one polarization per sub-correlator, both configurations provided 2048 spectral channels per sub-correlator and yielded a channel spacing of 48.8\,kHz (equivalent to 2.4\,km\,s$^{-1}$). 

Since  MG~J0414+0534 has quite strong continuum emission (0.71$\pm$0.02\,Jy, on average, the error being the standard deviation of the mean), we observed in double position switching mode \citep{ghosh02} to avoid problems related to residual standing waves in the baseline of the spectrum. A standard ON/OFF position-switched observation of 300s was performed on MG~J0414+0534, followed by a 40s ON/OFF observation on the strong continuum source 3C~120 (5.5$\pm$0.2\,Jy), which was used as a bandpass calibrator. The half power beam width (HPBW) was $\sim$0.7{\arcmin}$\times$\,0.9{\arcmin} and the pointing was found to be accurate within 10{\arcsec} in all observations. In order to obtain a precise flux calibration of our spectra, we also performed WAPP cross maps of the non-variable continuum source 3C~93, of the bandpass calibrator 3C~120, and of MG~J0414+0534. 

The data reduction was performed with the standard Arecibo Observatory (AO) IDL analysis package written by Phil Perillat using special routines developed by the AO staff. The individual ON/OFF scans on MG~J0414+0534 were processed to yield (ON-OFF)/OFF spectra, and these were divided by similar spectra for 3C~120 to obtain bandpass corrected spectra of MG~J0414+0534. The flux density of 3C~93, calculated using the K{\"u}hr's coefficients \citep{kuehr81}, was used to convert the resulting ratio spectra to Jy. The uncertainty of this flux calibration procedure is dominated by  the error on the flux density determined for 3C~93 and is estimated to be 7\%.
For each epoch, individual scans were inspected for quality and radio frequency interference (RFI) and co-added to produce a final spectrum. A polynomial baseline (typically of order 7--8) was then fitted to this spectrum and subtracted from it. Finally, we averaged the two polarizations. Due to a technical problem in one of the polarization channels of the June 2009 dataset, only a single polarization spectrum is reported for this epoch. The r.m.s. sensitivities reached in individual epochs ranged from 0.2 to 0.6 mJy per 2.4 km\,s$^{-1}$ wide channel (see Table~\ref{table:obs}). We measured the continuum flux density of MG~J0414+0534 from the calibrated cross maps.

%
\begin{table*}
\caption{Observational details}             
\label{table:obs}      
\centering 
\begin{tabular}{c l r c c c l}     
\hline\hline        
Epoch & Date           & Day No. & Receiver & On-source int. time & R.m.s.                          & Comments        \\
      &                &         &          & (minutes)           & (mJy per 2.4\,km\,s$^{-1}$ chan)  &                 \\ 
\hline                 
   1  & 2008 Oct 14-15 & 0       & C-high   &  50                 &  0.3   &                 \\ 
   2  & 2008 Nov 21-22 & 38      & C-high   &  55                 &  0.4   &                 \\ 
   3  & 2009 Jan 1-2   & 79      & C-high   &  50                 &  0.4   &                 \\ 
   4  & 2009 Feb 14-19 & 123     & C        &  195                &  0.2   &                 \\ 
   5  & 2009 Apr 4-5   & 172     & C        &  65                 &  0.4   &                 \\ 
   6  & 2009 May 16-17 & 214     & C        &  65                 &  0.5   &                 \\
   7  & 2009 Jun 27-28 & 256     & C        &  65                 &  0.6   & Single pol. spectrum \\
   8  & 2009 Aug 8-9   & 298     & C-high   &  65                 &  0.5   & dual board set up \\
   9  & 2009 Sep 28-30 & 349     & C        &  65                 &  0.4   &                   \\
  10  & 2009 Nov 12-13 & 394     & C-high   &  55                 &  0.4   & dual board set up \\
  11  & 2010 Jan 11-12 & 454     & C-high   &  45                 &  0.4   & dual board set up \\
\hline
\end{tabular}
\end{table*}

%

%
\begin{table}
\caption{The list of molecular transitions that were targeted.}            
\label{table:trans}      
\centering                          
\begin{tabular}{c c l @{}c}       
\hline\hline                 
 Band & Frequency        & Transitions   & Rest frequency \\
      & (GHz)            &               & (GHz)          \\    
\hline                                                   
   1  & 6.110            & H$_2$O $6_{16}-5_{23}$ & 22.235 \\
   2  & 6.515            & NH$_3$ (1, 1)         & 23.694 \\ 
      &                  & NH$_3$ (2, 2)         & 23.722 \\
   3  & 6.567            & OH $^{2}\Pi_{3/2} J=9/2, F=4-4$ & 23.817 \\
      &                  & OH $^{2}\Pi_{3/2} J=9/2, F=5-5$ & 23.826 \\
      &                  & NH$_3$ (3, 3)         & 23.870 \\
   4  & 6.878            & CH$_3$OH $4_{2}-4_{1}$ & 24.933 \\
      &                  & CH$_3$OH $6_{2}-6_{1}$ & 25.018 \\
      &                  & NH$_3$ (6, 6)         & 25.056 \\
      &                  & CH$_3$OH $7_{2}-7_{1}$ & 25.124 \\
\hline                                                    
\end{tabular}
\end{table}
%


\section{Results}\label{sect:res}

In the following, the quoted line velocities are defined w.r.t. the optical redshift of MG~J0414+0534, $z$\,=\,2.639 \citep{lawrence95}, using the optical velocity definition in the heliocentric frame. Isotropic line luminosities and upper limits have been calculated using:
\begin{equation}\label{eq:lum}
\frac{L_{\rm line}}{\rm L_{\odot}}=\frac{1}{m}\frac{0.001}{1+z}\frac{\nu_{\rm line}}{\rm [GHz]}\frac{\int{S\,dv}}{\rm [Jy\,km\,s^{-1}]}\frac{D_{\rm L}^2}{\rm [Mpc^2]},
\end{equation}
where $m$ is the lensing magnification, $z$ is the redshift of the background source, $\nu_{\rm line}$ is the rest frequency of the transition, $\int{S\,dv}$ is the integrated flux density, and $D_{\rm L}$ is the luminosity distance. The lensing magnification for MG~J0414+0534 is estimated to be $\sim$35 \citep{trotter00}. This value for the magnification is used under the assumption that the line emission is coincident with the radio continuum. If the line emission is not associated with the continuum, then the lensing magnification could be larger or smaller than 35. The luminosity distance of MG~J0414+0534 is 21,790\,Mpc.

The errors on the quantities derived from the continuum and the maser line emission have been calculated in the following way. The error on the continuum flux density was determined by using the calibration uncertainty. The errors on the integrated and peak line flux densities, and the line widths of the Gaussian profiles were determined by considering both the statistical uncertainties associated with the Gaussian fits and the uncertainties from the absolute flux calibration. Finally, we deduced the error on the flux densities and velocities of the line peak (i.\,e. the maximum of the H$_2$O spectrum) using the r.m.s of a single channel and the channel separation, respectively. 

\subsection{The tentative satellite line}\label{sect:satellite}
\begin{figure}
   \centering
   \includegraphics[angle=90,width=9cm]{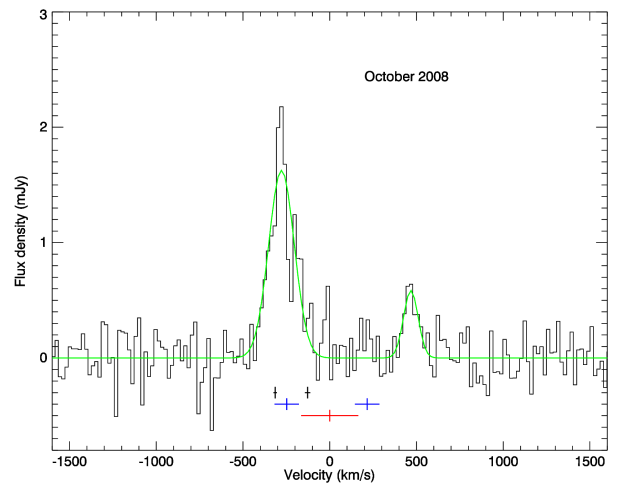}
      \caption{Water maser spectrum observed towards MG~J0414+0534 in October 2008 (black histogram). The fitted Gaussian profiles are overlaid (green line). Channel width is 19.2\,km\,s$^{-1}$. The root-mean-square (r.m.s.) noise level of the spectrum is 0.2\,mJy per channel. The velocity scale is relative to redshift 2.639 \citep{lawrence95} using the optical velocity definition in the heliocentric frame. The red cross marks the systemic velocity and the associated uncertainty (see Section~\ref{sect:red}). The blue and the black crosses indicate the peaks of the CO emission \citep{barvainis98} and the \ion{H}{i} absorption components \citep{moore99}, respectively, with their errors.}
         \label{fig:oct08_spec}
   \end{figure}

Our first Arecibo spectrum of MG~J0414+0534, taken in October 2008 (see Fig~\ref{fig:oct08_spec}), confidently confirms the presence of the water maser line that was detected in the discovery spectra obtained with Effelsberg and the EVLA \citep{impellizzeri08}. In addition, it shows a weak satellite emission feature, detected with a signal-to-noise ratio (SNR) of three, that is displaced by about +800\,km\,s$^{-1}$ from the main line. We fit simple Gaussian profiles to the maser features shown in Fig~\ref{fig:oct08_spec} and find that the main line has a central velocity of $-$278$\pm$5\,km\,s$^{-1}$ with a full width at half maximum (FWHM) of 174$\pm$5\,km\,s$^{-1}$. From the integrated flux density (0.30$\pm$0.03\,Jy\,km\,s$^{-1}$), using Eq.~\ref{eq:lum}, we derive for the main line an intrinsic (i.\,e. unlensed) H$_2$O isotropic luminosity of $\sim$26,000\,L$_{\odot}$  that makes the maser in MG~J0414+0534 the most luminous that is currently known.

The satellite line at +470$\pm$10\,km\,s$^{-1}$ has a FWHM of 100$\pm$10\,km\,s$^{-1}$ and is five times less luminous ($L_{\rm H_2O} \sim$5000\,L$_{\odot}$).  This feature could not be identified in the Effelsberg spectrum. Its peak flux density (0.6$\pm$0.2\,mJy) is comparable with  the r.m.s. noise level of the data (0.6\,mJy per 3.8\,km\,s$^{-1}$ channel; \citealt{impellizzeri08}). Smoothing the Effelsberg spectrum down to a channel width of 54\,km\,s$^{-1}$ (r.m.s $\sim$\,0.2\,mJy) still shows no significant emission around +470\,km\,s$^{-1}$. The velocity of the satellite line was not covered by the bandwidth of our discovery EVLA spectrum. Surprisingly, this emission line feature was not detected again after October 2008 (see Fig~\ref{fig:spectra}). In February 2009 we performed deeper observations aimed at confirming the presence of this feature. No emission line other than the main one at about $-$300\,km\,s$^{-1}$ was detected above a 3$\sigma$ noise level of 0.3\,mJy per 19.2\,km\,s$^{-1}$ channel. However, a weak feature is seen in the spectrum  at the velocity of about +490\,km\,s$^{-1}$ (see Fig.\ref{fig:fit4}, lower panel). The satellite line remains undetected also in the spectrum produced by averaging all of the epochs with the same weights (Fig~\ref{fig:fit4}, upper panel). Nonetheless, we note that the range between 200 and 500\,km\,s$^{-1}$ looks spiky and that, interestingly, one of these spikes is at the position of the satellite line. Averaging the spectra using different weights (e.\,g. 1/r.m.s$^2$ or the integration time) does not change the shape of the resulting spectrum. This may indicate that many weak lines are present in the range 200--500\,km\,s$^{-1}$ and that in October 2008 we saw one of these lines flaring.   

\subsection{Structure of the main line}\label{sect:fit4}
\begin{figure}
   \centering
   \includegraphics[angle=-90,width=9cm]{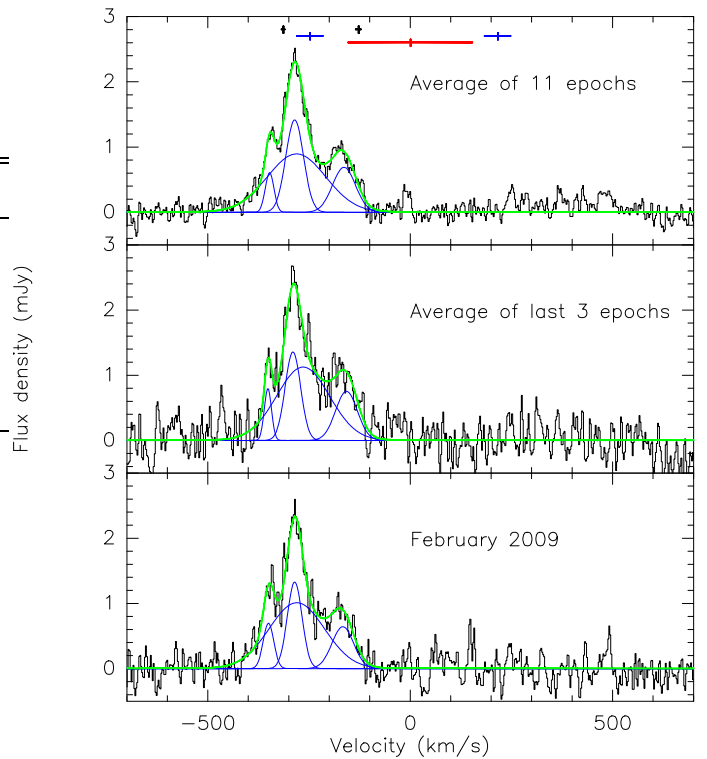}
      \caption{{\it Lower panel}: Water water maser spectrum of MG~J0414+0534 observed in February 2009. {\it Middle panel}: Average of the last three epochs (September and November 2009 and January 2010) obtained using equal weights. {\it Upper panel}: Final spectrum produced by averaging all the epochs with the same weight. Individual Gaussian profiles fitted to the spectra are overlaid in blue, while the resulting profile is displayed in green. The red cross marks the systemic velocity and the associated uncertainty (see Section~\ref{sect:red}). The blue and the black crosses indicate the peaks of the CO emission \citep{barvainis98} and the \ion{H}{i} absorption components \citep{moore99}, respectively, with their errors . Channel spacing is 2.4\,km\,s$^{-1}$. The r.m.s noise level is 0.2\,mJy per channel in the spectra of the lower and middle panels and 0.1\,mJy per channel in the upper panel.}
         \label{fig:fit4}
   \end{figure}

The high SNR of the February 2009 spectrum ($\sim$13; see Fig~\ref{fig:fit4}, lower panel) reveals that the main line has a complex profile that is likely the result of the blending of many components. When we fit the line profile with multiple Gaussians, the best fit is obtained using four components. Due to the lower SNR of the spectra, it is impossible to perform the same analysis for the other epochs. However, the four Gaussian components well describe the average profile of the main line (Fig~\ref{fig:fit4}, upper panel), implying that they must be present in most of the epochs. In order to inspect the variability of the individual velocity features, we produced a spectrum by averaging with equal weights the last three epochs of the monitoring (September and November 2009 and January 2010). The resulting spectrum (Fig~\ref{fig:fit4}, middle panel) has an r.m.s comparable with that of the February 2009 observation. The mean time separation between the two spectra is 276 days. Table~\ref{table:gauss} summarizes the properties of the Gaussian profiles fitted to these spectra (central velocity, FWHM, and integrated flux density) and the intrinsic, i.\,e. unlensed isotropic H$_2$O luminosity. Comparing the Gaussian peak velocities, we find that the velocity of components I and II did not change, while the velocities of components III and IV have marginally increased by +15$\pm$3\,km\,s$^{-1}$ and +10$\pm$3\,km\,s$^{-1}$, respectively. However, these weaker features can be identified in only two of the eleven spectra from individual epochs. It is therefore possible that the change in the peak velocities of these components is due to a change in the line profile rather than to an actual motion of the gas.

%
\begin{table}
\caption{Parameters of the Gaussian profiles fitted to the water maser line in the spectra of February 2009 (epoch~4) and the average of the last three epochs (epochs 9, 10, and 11; see also Fig.~\ref{fig:fit4}).}            
\label{table:gauss}      
\centering                          
\begin{tabular}{c c c c c r}       
\hline\hline                 
 Comp. & Epoch     & Velocity       & FWHM          & Int. flux density   & Lum.          \\
       &           & (km\,s$^{-1}$) & (km\,s$^{-1}$) & (mJy\,km\,s$^{-1}$) & (L$_{\odot}$) \\    
\hline                                                  
   I   &  4        & -350$\pm$2     & 31$\pm$2      & 23$\pm$12           & 2000         \\
       & 9, 10, 11 & -351$\pm$2     & 21$\pm$2      & 17$\pm$5            & 1500         \\
   II  &  4        & -285$\pm$2     & 43$\pm$2      & 60$\pm$12           & 5100         \\
       & 9, 10, 11 & -290$\pm$2     & 45$\pm$2      & 65$\pm$5            & 5600         \\
   III &  4        & -280$\pm$2     & 161$\pm$2     & 173$\pm$12          & 14800        \\
       & 9, 10, 11 & -265$\pm$2     & 154$\pm$2     & 184$\pm$5           & 15800        \\
   IV  &  4        & -167$\pm$2     & 63$\pm$2      & 43$\pm$12           & 3700         \\
       & 9, 10, 11 & -157$\pm$2     & 63$\pm$2      & 51$\pm$5            & 4400         \\
\hline                                                    
\end{tabular}
\end{table}
%

\subsection{Monitoring}

\begin{figure*}
   \centering
   \includegraphics[scale=0.8]{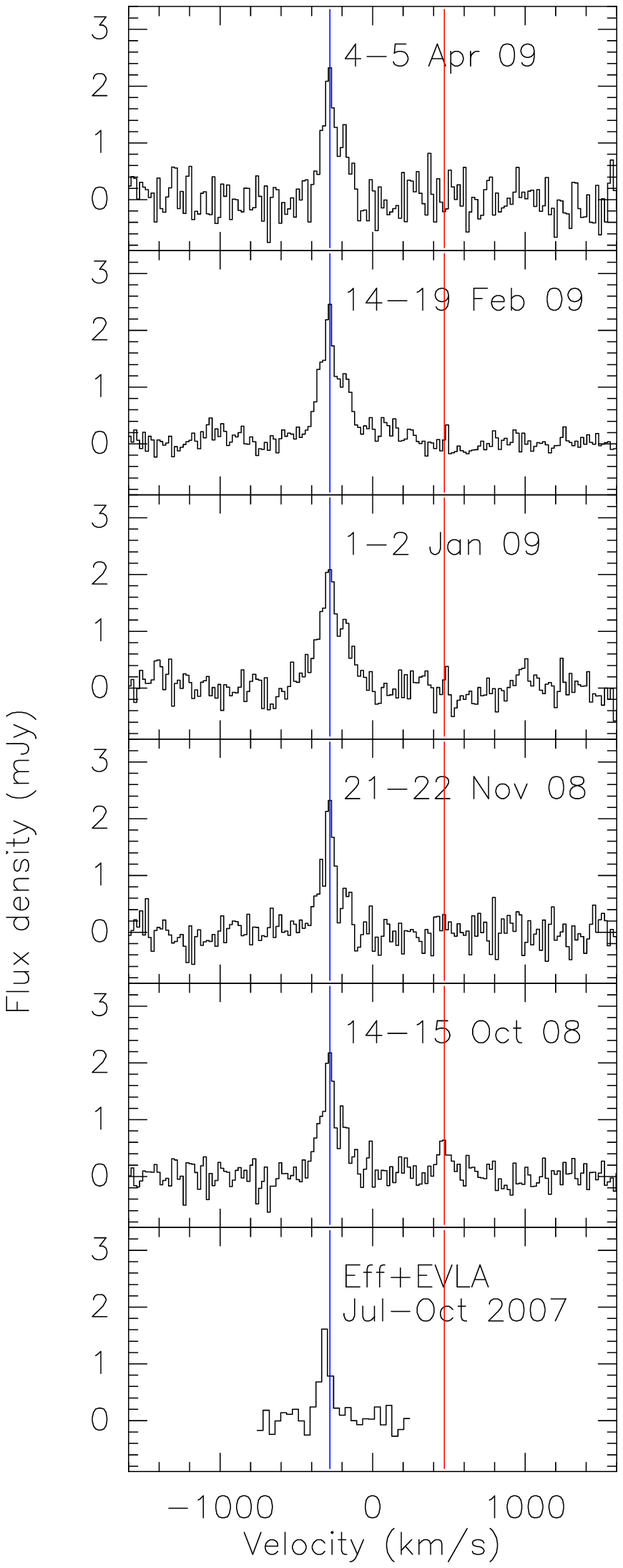}
   \hspace{0.5cm}
   \includegraphics[scale=0.8]{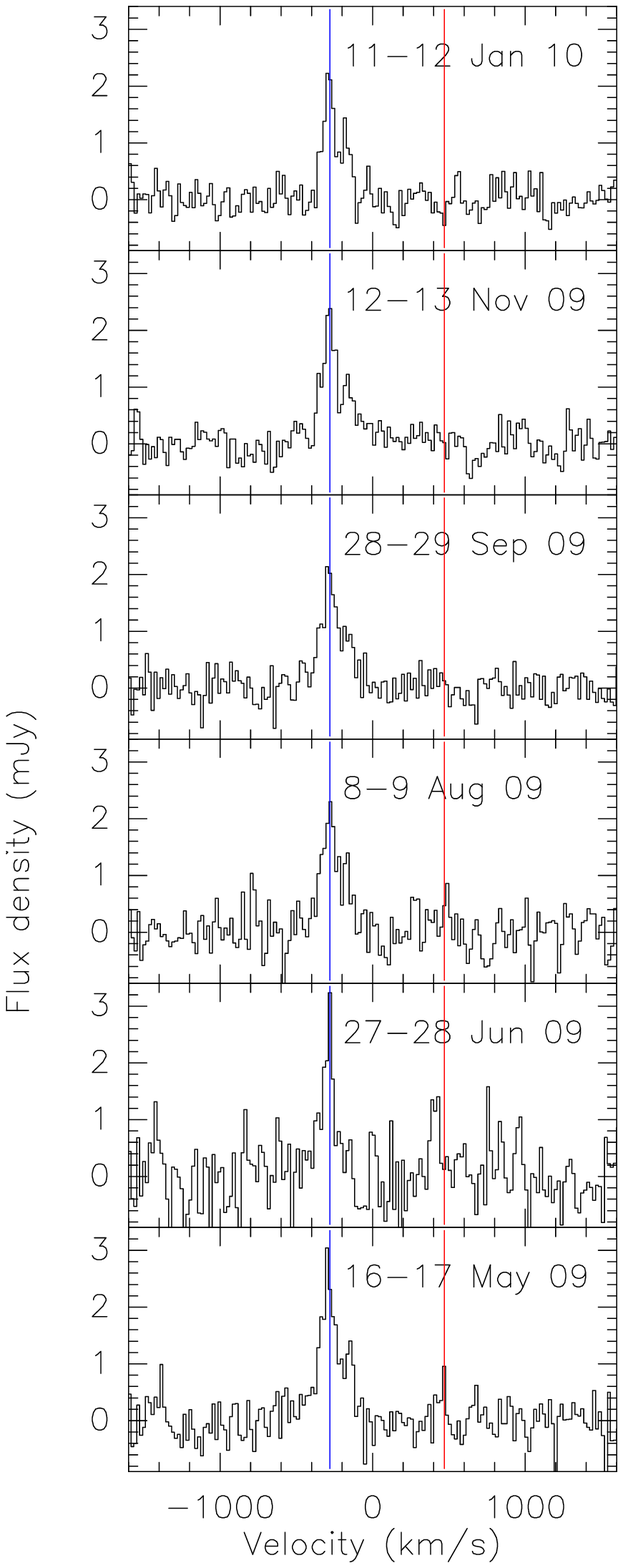}
   \caption{Water maser spectra observed towards MG~J0414+0534 between July 2007 and January 2010. The first spectrum (bottom left corner) is the combined Effelsberg and EVLA spectrum (channel spacing 38.4\,km\,s$^{-1}$) obtained observing 14+12\,hrs on-source between July and October 2007 (for details, see \citealt{impellizzeri08}). The other spectra have been taken with Arecibo. The spectra shown here have been smoothed down to a resolution of 19.2\,km\,s$^{-1}$. The blue and red vertical lines indicate the peak velocities of the main and satellite maser features, respectively, as measured in the October 2008 spectrum.}
              \label{fig:spectra}
    \end{figure*}

\begin{figure*}
   \centering
   \includegraphics[angle=90,width=9.1cm]{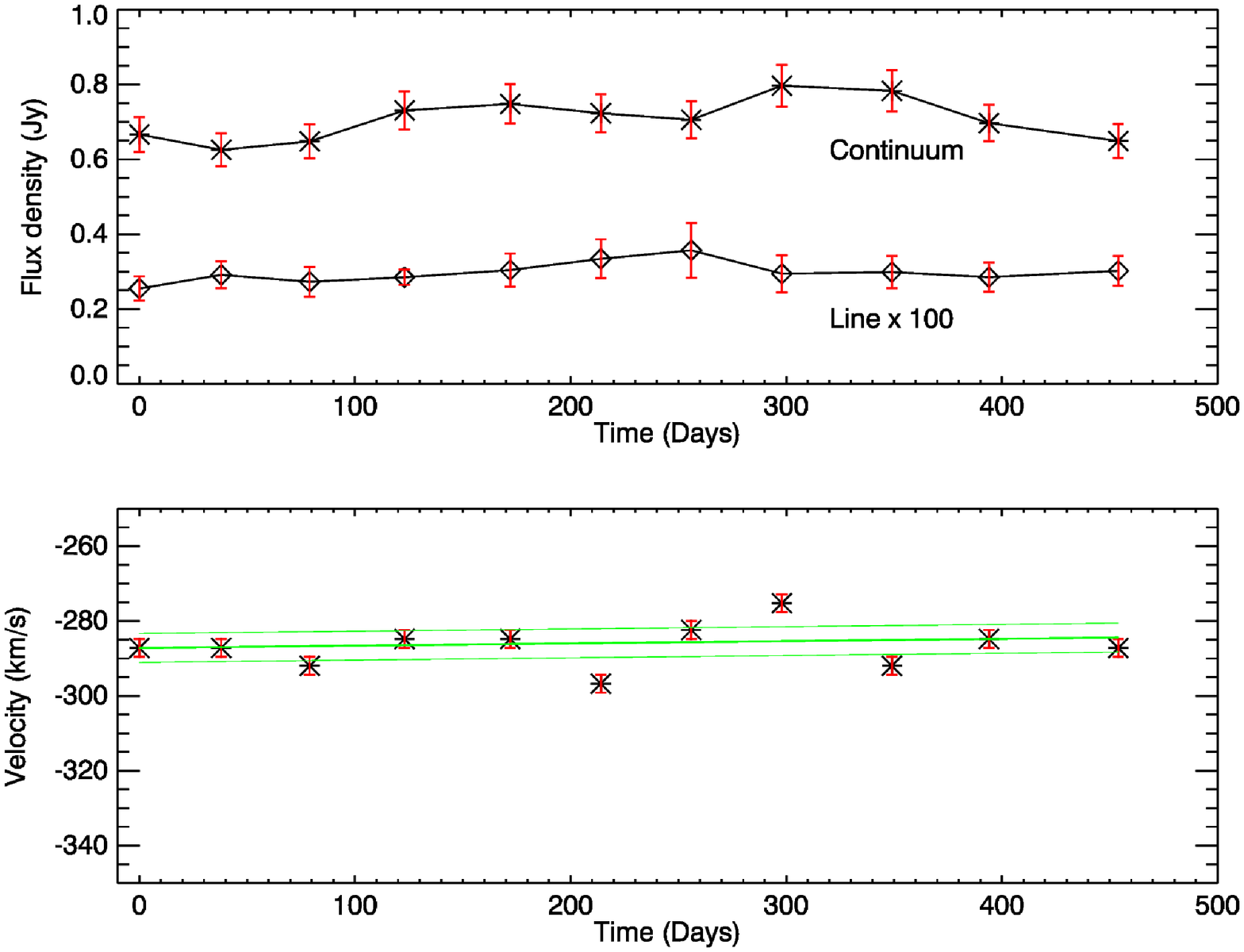}
   \includegraphics[angle=90,width=9.1cm]{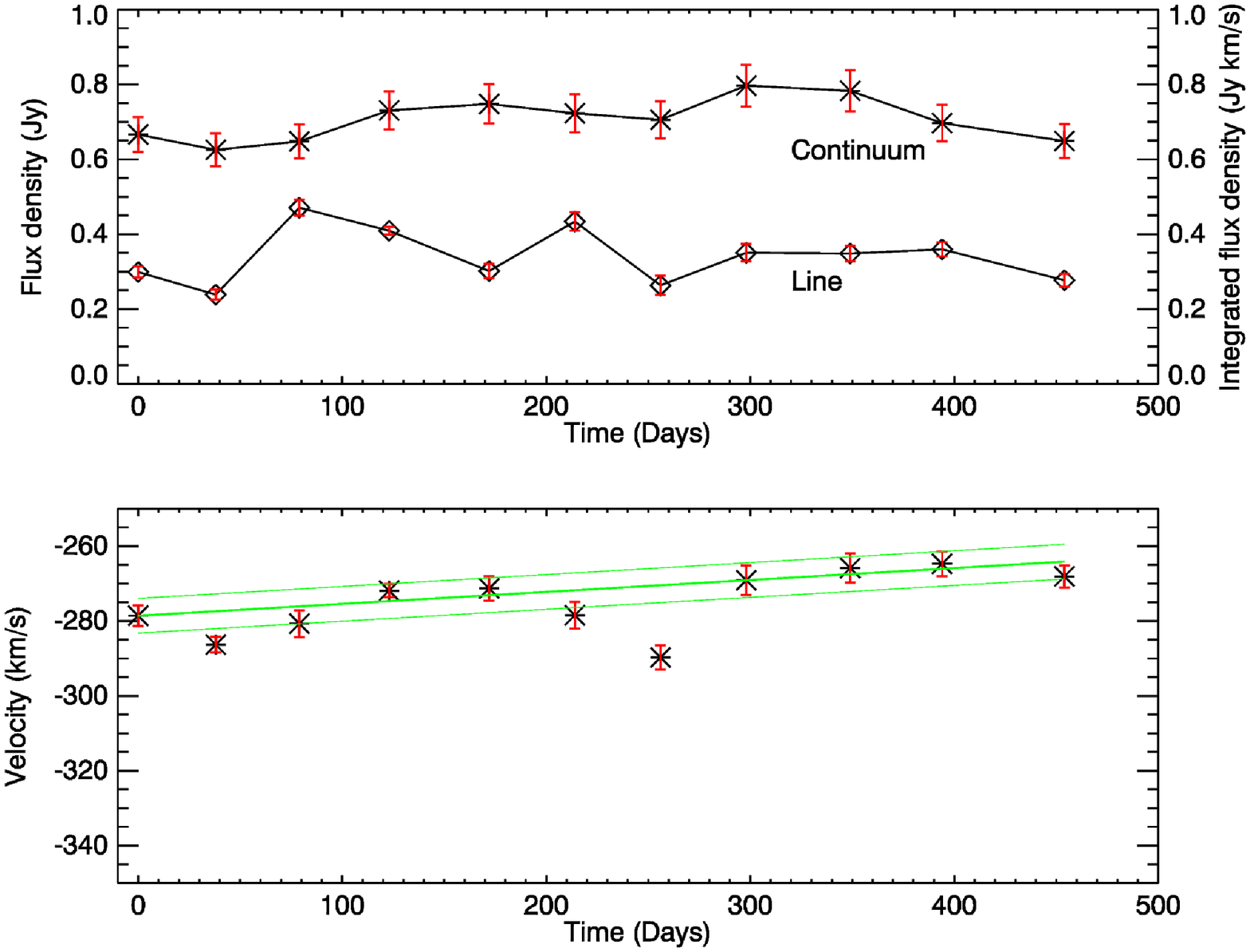}
      \caption{{\it Left panels}: Continuum and line peak flux density vs. time (up) and peak velocity vs. time (bottom) for the spectra with a channel spacing of 2.4\,km\,s$^{-1}$. The peak flux density of the line has been multiplied by a factor of 100 to allow the comparison with the continuum flux density within the same plot. The error bars represent the r.m.s. noise levels per channel of the spectra and the width of the channels, respectively. {\it Right panels}: Continuum and integrated flux density vs. time (up) and peak velocity vs. time (bottom) derived from Gaussian fitting. The error bars indicate the formal errors obtained from the fits. Best fit lines and mean absolute deviations are shown in green. Day 0 is 2008 October 14-15}
              \label{fig:monitoring}
    \end{figure*}

The results of our continuum and line monitoring are displayed in Figs.~\ref{fig:spectra} and~\ref{fig:monitoring}. Fig.~\ref{fig:spectra} shows the sequence of spectra observed towards MG~J0414+0534 from July 2007 to January 2010. In addition to the Arecibo spectra, we also show the combined Effelsberg and EVLA spectrum. This spectrum is the combination of 14 hours of observations with the Effelsberg radio telescope between July and September 2007, and 12 hours of observations with the EVLA in October 2007 (for details see \citealt{impellizzeri08}). Although the line profile does change slightly from epoch to epoch, the overall appearance of the main H$_2$O emission feature remains stable during the period of the Arecibo observations. A significant change in the line profile seems instead to have occurred between the Effelsberg and EVLA observations and the first epoch of the Arecibo monitoring campaign. The line appears to be much broader in the first Arecibo spectrum w.r.t. the previous observations. This is confirmed by a comparison between the Gaussian fit parameters of the lines. Fitting a single Gaussian profile to the combined Effelsberg and EVLA spectrum, we obtain a FWHM of 78$\pm$4\,km\,s$^{-1}$ which is about half the line width measured in the Arecibo spectrum of October 2008 (see Section~\ref{sect:satellite}). This line broadening is responsible for the larger intrinsic isotropic luminosity we measure (26,000\,L$_{\odot}$) w.r.t. that reported by \citet{impellizzeri08} (10,000\,L$_{\odot}$). The line velocity is also different. Correcting for the small shift in the reference frequency used in the Arecibo observations (6110.0\,MHz) w.r.t that used at Effelsberg and the EVLA (6110.2\,MHz) the peak of the Gaussian in the combined spectrum is at -312$\pm$4\,km\,s$^{-1}$. Hence, the line is redshifted by 34$\pm$6\,km\,s$^{-1}$ in the Arecibo spectrum. Differences in the baseline calibration of the datasets, though possibly accounting in part for the different line widths, are not sufficient to explain these discrepancies. The most plausible interpretation is that there was a real change in the line profile. Looking at the spectra in Fig.~\ref{fig:spectra} it seems that the most redshifted component seen in Fig.~\ref{fig:fit4} (component 4 in Table~\ref{table:gauss}) was not present at the time of the Effelsberg and EVLA observations.     

In Fig.~\ref{fig:monitoring} (left panels), we plot the 6\,GHz continuum flux density of MG~J0414+0534 together with the peak flux density and the peak velocity of the line as a function of time. In the right panels instead, the continuum flux density is displayed together with the integrated flux density and the Gaussian peak velocity of the line as a function of time. The integrated flux density and the Gaussian peak velocities have been derived by fitting a single Gaussian profile to the broad maser feature. Absolute deviations of the continuum flux from the mean are, on average, comparable with the flux calibration uncertainty (7\% see Section~\ref{sect:obs}). The 6\,GHz continuum flux density of MG~J0414+0534 thus remained nearly constant for the duration of the whole monitoring period, with an average flux density of 0.71$\pm$0.02\,Jy. The line peak flux density is also surprisingly stable throughout the period of the observations. Small fluctuations are not exceeding the limits of uncertainty (between 10\% and 50\%). The integrated flux density instead, shows significant variations from epoch to epoch that, however, do not follow a definite trend. The behaviour of the integrated flux density reflects the variation of the width of the Gaussian profile, whose FWHM fluctuates between $\sim$100 and $\sim$240\,km\,s$^{-1}$ during the monitoring period. This variation is likely the result of flux variability among individual velocity components (see Section~\ref{sect:fit4}). 

We fit a linear function to the line and Gaussian peak velocities. In both cases, the $\chi^2$ values calculated from the fits are quite high, indicating that, most likely, if there is a systematic acceleration, this is not constant. Nevertheless, assuming that a straight line is the correct model for the data, we can calculate the accelerations using a least absolute deviation method, which is less sensitive to outlying data w.r.t the $\chi^2$ minimization.  The best fit lines and the mean absolute deviations are shown in Fig.~\ref{fig:monitoring} (lower panels). We find that the line peak velocity is constant within the limit of the uncertainty associated with the peak identification (i.\,e. the channel width, 2.4\,km\,s$^{-1}$). The line velocity derived from Gaussian fits instead, is increasing by $\sim$12\,km\,s$^{-1}$\,yr$^{-1}$. However, since the Gaussian fit is sensitive to the whole profile, this trend may be due to fluctuations in the relative intensities of the individual velocity components rather than to a real acceleration of the masing clouds. Furthermore, drifting maser lines, as those observed in edge on accretion disks, typically have line widths of 1-4\,km\,s$^{-1}$ (e.\,g. NGC~4258; \citealt{humphreys08}). Velocity drifts of broad (FWHM\,$\sim$100\,km\,s$^{-1}$) maser features have never been observed so far. Thus, we treat this result with caution and do not use it in our discussion.
  
%
 
\subsection{Upper limits on the other molecular transitions}

In three occasions, August 2009, November 2009, and January 2010, we took advantage of the WAPP dual board mode to search for molecular emission lines from NH$_3$, OH, and CH$_3$OH (see Table~\ref{table:trans}) towards MG~J0414+0534.  No emission line was detected in the individual spectra nor in the average spectrum above a 5\,$\sigma$ noise level of 1.5 and 2.0\,mJy per 2.4\,km\,s$^{-1}$ channel, for bands 2 and 3, respectively. Using Eq.~\ref{eq:lum} and considering rectangular lines of width 2.4\,km\,s$^{-1}$, this yields upper limits on the isotropic luminosities of the ammonia inversion lines (1, 1), (2, 2), and (3, 3) of $\sim$330\,L$_{\odot}$ (for the first two transitions) and $\sim$440\,L$_{\odot}$. The luminosity of the two OH lines must be $<$\,440\,L$_{\odot}$. Unfortunately, the frequency band centered at 6.9\,GHz, where the NH$_3$ (6, 6) and the excited CH$_3$OH transition frequencies fall, is severely affected by RFI. Interferences are present in about 60\% of the band making any line identification impossible.


\section{Discussion}\label{sect:disc}

We have monitored the radio continuum and maser emission in MG~J0414+0534 for $\sim$15 months at $\sim$6 week intervals and found that both are surprisingly stable. The continuum and the line peak flux density were found to be constant throughout the periods of observations. The integrated flux density instead displays significant changes from epoch to epoch that are likely the result of changes in the individual velocity components. From the analysis of the 11 epochs of the monitoring, we can place an upper limit on the velocity drift of the most prominent line peak (component II in Table~\ref{table:gauss}) of 2\,km\,s$^{-1}$\,yr$^{-1}$. We tentatively detected a weaker satellite line at +470\,km\,s$^{-1}$ in October 2008 that, however, was not confirmed by the spectra of the other epochs, nor by our most sensitive spectrum obtained by averaging all the epochs. In the next sections we examine the possible scenarios for the origin of the maser in the light of these results. 
 
\subsection{The redshift of MG~J0414+0534}\label{sect:red}

For the discussion presented here it is of fundamental importance to assess the true redshift of MG~J0414+0534 and the accuracy of the corresponding systemic velocity. \citet{lawrence95} derived a redshift of 2.639$\pm$0.002 from the broad H$\alpha$ emission line identified in the infrared spectrum of the quasar. CO (3--2) emission was detected toward MG~J0414+0534 by \citet{barvainis98} and is centered at the H$_{\alpha}$ redshift, while \ion{H}{i} is seen in absorption, blueshifted by $\sim$200\,km\,s$^{-1}$ w.r.t. the H$_{\alpha}$ and CO emission lines \citep{moore99}. The \ion{H}{i} line consists of two absorption components, one at $z=2.6376\pm0.0002$ and one at $z=2.6353\pm0.0001$, with the most prominent and blueshifted one approximately coincident with the peak of the H$_2$O emission (Figs.~\ref{fig:oct08_spec} and \ref{fig:fit4}). 

The discrepancy between the redshift of the H$_{\alpha}$ emission and of the \ion{H}{i} absorption centroid is not surprising. In fact, previous studies on various types of galaxies indicate that systemic velocities derived from optical emission lines can be biased by motions of the emitting gas and obscuration of the back side of the galaxy \citep{mirabel84,morganti01}. More remarkable is the difference between the redshift of the CO and the \ion{H}{i} lines, given that CO traces the large scale galaxy structure and should be free of outflow/infall problems. According to \citet{moore99}, in the case of MG~J0414+0534, this offset  might be due to i) the \ion{H}{i} absorption occurring against an extended jet component and not towards the nucleus or ii) the \ion{H}{i} is absorbing the active nucleus and CO emission  has a different spatial distribution and is affected differently by gravitational lensing. In the following, we assume that the optical/CO redshift, $z=2.639$, is the most reliable redshift for MG~J0414+0534. For the sake of completeness, we will also discuss the possibility that the redshift of MG~J0414+0534 is the one derived from \ion{H}{i} absorption. The uncertainty in the optical redshift determination corresponds to a large uncertainty in the definition of the systemic velocity ($\pm$165\,km\,$^{-1}$). Accordingly, the main maser line is mostly blueshifted w.r.t the systemic velocity of MG~J0414+0534, although part of the emission may possibly be considered systemic (see Fig.~\ref{fig:fit4} and Table~\ref{table:gauss}). 

\subsection{Origin of the H$_2$O emission}
\subsubsection{Jet-maser scenario} 
Our initial hypothesis, based on the absence of systemic and redshifted components in the Effelsberg and EVLA spectra and on the wide line profile, was that the emission is associated with the prominent relativistic jets of the quasar \citep{impellizzeri08}. Part of our results are indeed consistent with this interpretation. First of all, even when the maser line profile is resolved into multiple velocity components, individual emission features have line widths between 30 and 160\,km\,s$^{-1}$ that resemble those of known H$_2$O masers associated with radio jets (e.\,g. Mrk~348; \citealt{peck03}). Our non-detection of a radial acceleration of the main maser peak is also compatible with the jet-maser scenario. 

Adopting the hypothesis that the maser in MG~J0414+0534 is associated with the jet(s), the H$_2$O emission may arise from the shocked region at the interface between the relativistic jet material and an encroaching molecular cloud, as is believed to be the case for the masers in Mrk~348 \citep{peck03} and part of the emission in NGC~1068 \citep{galli01}. Alternatively, it could also be the result of the amplification of the radio continuum of the jet by foreground molecular clouds along the line of sight (as in NGC~1052; \citealt{sawada08}). In this framework, the maser and continuum intensities in MG~J0414+0534 would then be expected to show a similar behaviour to that in the aforementioned cases. For Mrk~348, strong variation of both maser and continuum flux densities are reported, with a close temporal correlation between them \citep{peck03}. The peak flux density of the jet-maser component in NGC~1068 is also variable, although the variability is not outstanding \citep{galli01}. For the third jet-maser case, that of NGC~1052, the variability is mainly caused by changes in the line profile \citep{braatz03}. The extreme stability of the main line peak and continuum flux density in MG~J0414+0534 resulting from our study,  seems to exclude a jet-maser scenario similar to that in Mrk~348 and NGC~1068, while the reported significant variations in the line profile of our target, may hint at similarities with the case of NGC~1052. We note, however, that the number of sources in which the maser emission is confidently associated with the jet(s) is very low and that more of these masers should be studied in detail in order to investigate the properties of these kind of sources.

The possibility that the \ion{H}{i} absorption occurs against a jet component and not against the core (see Section~\ref{sect:red}) is interesting and might favour the jet-maser scenario. Indeed, the most blueshifted \ion{H}{i} component is displaced by only (29$\pm$8)\,km\,s$^{-1}$ from the peak of the H$_2$O emission suggesting that the same gas structure that is absorbing the continuum radiation from the jet may host the molecular clouds that produce the maser emission. 

\subsubsection{Disk-maser scenario} 

The presence of highly red- and blueshifted emission features, symmetrically bracketing the systemic velocity, and emission lines close to the systemic velocity (typically referred to as `triple-peak profile') is a spectroscopic signature of masers that are associated with edge-on accretion disks (e.\,g. NGC4258; \citealt{n4258miyo}). Within this frame, the tentative detection of the redshifted line at +470\,km\,s$^{-1}$ in October 2008, may be seen as an element in favour of the accretion disk scenario. According to the standard model, we expect the high-velocity emission to arise from near the mid-line of the disk, defined as the diameter of the disk perpendicular to the line of sight, while the maser emission at the systemic velocity should originate on the near side of the disk. Therefore, the predicted radial accelerations of the high-velocity features are much smaller than those of the lines near the systemic velocity. The velocity drifts measured for the high-velocity maser lines in NGC~4258, for example, are in the range -0.7 to +0.7\,km\,s$^{-1}$\,yr$^{-1}$ \citep{humphreys08}. Our upper limit on the radial acceleration of the blueshifted maser emission in MG~J0414+0534, 2\,km\,s$^{-1}$\,yr$^{-1}$, cannot rule out such accelerations. 

If the main maser line and the satellite line at +470\,km\,s$^{-1}$ can be considered as the blueshifted and redshifted lines from the tangentially seen part of an edge-on accretion disk in Keplerian rotation, then the radius at which emission originates is given by $R= GM_{\rm BH}V_{\rm R}^{-2}$, where $G$ is the gravitational constant, $M_{\rm BH}$ is the black hole mass, and $V_{\rm R}$ is the rotational velocity at radius $R$. From the difference between the line of sight velocities of the main and satellite maser lines ($V_{\rm obs}$), we obtain $V_{\rm R} = V_{\rm obs} \cdot \sin (i)^{-1}$ $\sim 370 \cdot \sin (i)^{-1}$\,km\,s$^{-1}$. Adopting the black hole mass of $M_{\rm BH}=10^{9.0}$\,M$_{\odot}$ calculated by \citet{pooley07} for MG~J0414+0534, and assuming an edge-on orientation ($i$ = 90{\degr}) for the accretion disk\footnote{Since accretion disks that provide enough amplification paths for maser action have inclinations that differ less than 10{\degr} from an edge-on orientation (see e.\,g., \citealt{kuo2011}), the values for the rotation velocity, and hence, the radius and gas density of the disk, should not be very different from the one calculated assuming $i$ = 90{\degr}.}, we get a radius of $R \sim$ 30\,pc. This value is fairly large compared to the radii at which maser emission is found in the accretion disks of nearby radio quiet AGN (typically, 0.1 to 1 pc). We should keep in mind however, that MG~J0414+0534 is a radio loud quasar, while known disk-maser hosts are mainly radio quiet Seyfert or LINER galaxies with a mass of the nuclear engine that is two orders of magnitude lower ($\sim 10^7$\,M$_{\odot}$; \citealt{kuo2011}). 

In order to understand if the physical conditions of the gas at 30\,pc from a 10$^9$M\,$_{\odot}$ black hole are suitable to provide water maser emission, we calculate the density of the gas necessary to reach stability against tidal disruption, in a spherical clump at a radius $R$ from the central engine that is rotating at a velocity $V_{\rm R}$ (see \citealt{tarchi07} and references therein). For MG~J0414+0534 we obtain that the density of H$_2$ molecules at a radius of 30\,pc from the nuclear engine would need to be $\ga 2 \times 10^5$cm$^{-3}$. Such a density is far from the density at which the H$_2$O level population thermalize (e.\,g. \citealt{kylafis91}) and, hence, the conditions of the gas do not preclude the production of water maser emission. Therefore, the identification of the tentative line at about +470\,km\,s$^{-1}$\,yr$^{-1}$ as the redshifted feature of the characteristic disk-maser spectrum is physically plausible, thus making the disk-maser picture a viable option.   

If we assume that the atomic gas is absorbing the radio continuum emission from the core (see Section~\ref{sect:red}) and that MG~J0414+0534 is at the redshift of the \ion{H}{i} absorption centroid (2.6365; \citealt{moore99}), the value of the maser disk radius calculated above changes substantially. In fact, in this case the main line lies at the systemic velocity of the quasar and the inferred rotational velocity and radius are $\sim$750\,km\,s$^{-1}$ and $\sim$7\,pc, respectively. For such a disk to be stable the density of H$_2$ molecules at a radius of 7\,pc from the nuclear engine would need to be $\ga 10^7$cm$^{-3}$, a density that is still compatible with the production of H$_2$O maser emission in the 6$_{16}$--5$_{23}$ transition (e.\,g. \citealt{kylafis91}). In the hypothesis that the maser emission originates on the near side of the disk, the velocity drift is given by $V^2_{\rm R} R^{-1}$. Assuming that the radius at which the systemic and high-velocity lines arise is the same, we obtain a velocity drift of $\sim$0.8\,km\,s$^{-1}$\,yr$^{-1}$. A longer monitoring period (at least 4 or 5 years) and/or a higher spectral resolution would be necessary to detect such a small velocity drift and test this hypothesis. 

Therefore, although the type~1 optical spectrum and the relatively low column density derived from X-ray observations (N$_{\rm H}$\,$\sim 5 \times 10^{22}$\,cm$^{-2}$; \citealt{chartas02}) indicate that the disk might not be favourably oriented to produce detectable water maser emission, we cannot exclude this possibility on the basis of our single-dish data alone.


\section{Conclusions}\label{sect:con}
The redshifted 6\,GHz radio continuum and H$_2$O maser emission in the type~1 quasar MG~J0414+0534 at $z=2.639$ have been monitored with the 300-m Arecibo telescope for $\sim$15 months, in order to help shedding light on the origin of the most distant water maser found to date. 

We have confirmed the H$_2$O detection reported by \citet{impellizzeri08} at high signal-to-noise levels and have found that the line profile can be resolved into a complex of features with line widths between 30 and 160\,km\,s$^{-1}$. A redshifted line was tentatively detected in October 2008 at a velocity of +470\,km\,s$^{-1}$. The total intrinsic (i.\,e. unlensed) H$_2$O isotropic luminosity is $\sim$30,000\,L$_{\odot}$ making the maser in MG~J0414+0534 the most luminous ever discovered. The overall appearance of the main maser feature, as well as the flux density of the most prominent peak, are surprisingly stable throughout the period of the observations, although the integrated flux density shows significant variations on monthly time scales, possibly hinting at changes in the individual velocity components. The continuum flux density is also quite stable from epoch to epoch. The velocity of the strongest line peak is constant within the uncertainty, thus providing an upper limit on the velocity drift of 2\,km\,s$^{-1}$\,yr$^{-1}$.

The large line widths of the individual velocity components of the H$_2$O maser feature and the lack of an evident triple-peak profile favour an association of the maser with the relativistic jet(s) of the quasar. The type~1 nature of the AGN in MG~J0414+0534 further reinforces this interpretation. However, the remarkable stability of the continuum and the line emission is partly in contrast with this picture. Furthermore, the tentative detection of the redshifted feature in the October 2008 spectrum is compatible with the disk-maser hypothesis. 

While providing useful clues to determine the nature of the maser in MG~J0414+0534, our single-dish data alone are presently insufficient to confidently exclude either one of the two scenarios, jet vs. accretion disk. VLBI observations and longer time-scale single-dish monitoring will be essential to unveil the origin of the H$_2$O maser in this intriguing object.

\begin{acknowledgements}
P.C. and V.I. wish to thank the operators at the 300--m telescope for their hospitality during the observing runs. We are indebted to C. Salter and T. Ghosh for their invaluable assistance during observing preparation  and data reduction and to H. Hernandez for the careful scheduling of this long monitoring. We are also grateful to the anonymous referee for his/her useful suggestions. P.C. wish to thank A. Tarchi for critically reading the manuscript. O.W. is funded by the Emmy-Noether-Programme of the `Deutsche Forschungsgemeinschaft', reference Wu 588/1-1.
\end{acknowledgements}

\bibliographystyle{aa} 
\bibliography{castangia2011} 

\end{document}